\documentclass[12pt,a4paper,keywords,prb]{revtex4}

\usepackage{graphicx}
\usepackage{amsmath}
\usepackage{bm}

\begin{document}

\title{Pinning of stripes by local structural distortions in cuprate high-T$_c$ superconductors}
\author{U. Tricoli$^{1,2}$ and B. M. Andersen$^2$}
\affiliation{$^1$Dipartimento di Fisica dell'Universit\'{a} di Bologna and INFN, via Irnerio 46, 40126 Bologna, Italy\\
$^2$Niels Bohr Institute, University of Copenhagen, Universitetsparken 5, DK-2100 Copenhagen,
Denmark}

\begin{abstract}

We study the spin-density wave (stripe) instability in lattices with mixed low-temperature orthorhombic (LTO) and low-temperature tetragonal (LTT) crystal symmetry. Within an explicit mean-field model it is shown how local LTT regions act as pinning centers for static stripe formation. We calculate the modulations in the local density of states near these local stripe regions and find that mainly the coherence peaks and the van Hove singularity (VHS) are spatially modulated. Lastly, we use the real-space approach to simulate recent tunneling data in the overdoped regime where the VHS has been detected by utilizing local normal state regions. 

\end{abstract}

\date{\today}
\maketitle


\section{Introduction}

By now it has become evident that stripes play an important role in the understanding of cuprate high-T$_c$ superconductors.\cite{kivelson,tranquada} This is certainly true for materials where the underlying lattice breaks $C_4$ symmetry such as the LTT phase of e.g. La$_{1.6-x}$Nd$_{0.4}$Sr$_x$CuO$_4$ or La$_{2-x}$Ba$_x$CuO$_4$ (LBCO).\cite{tranquada95,fujita,hucker} The situation is, however, complicated by the presence of disorder and grains which can pin stripes and restore global $C_4$ symmetry. In intrinsically disordered materials such as La$_{2-x}$Sr$_x$CuO$_4$ (LSCO) experiments have shown that a static magnetic signal persists well into the superconducting dome and vanishes around a doping level of $\sim 13\%$.\cite{kivelson,julien}

Theoretically, the stripe disorder has been studied within a variety of models.\cite{alvarez,robertson,delmaestro,BMAndersen:2007,atkinson,syljuaasen,Schmid10} In particular, a mean-field unrestricted Hartree-Fock study including potential disorder to model dopant and substitutional impurities demonstrated how disorder slows-down the spin fluctuations, which eventually freeze out and form a disordered low-temperature static stripe phase.\cite{andersen10} The same picture applies in the presence of vortices generated by a magnetic field.\cite{zhu,Schmid10,AndersenJPCS}   

Another candidate for pinning centers of stripe fluctuations is local LTT regions embedded in the LTO matrix. This is relevant in, for example, LSCO where various experimental studies have detected local LTT distortions caused by disorder, dopants, and grain boundaries.\cite{bianconi,saini,bozin,horibe,codero} The LTT phase is clearly also relevant for LBCO where it exists as the preferred bulk crystal structure around 1/8 doping.\cite{hucker} In this bulk LTT phase it is well-known from experimental studies that electron-lattice coupling is important for pinning of stripes.\cite{crawford,buchner,tranquada1996} In LSCO the static spin signal also peaks close to 1/8 doping\cite{julien} possibly due to disordered local LTT centers dominating the pinning near that doping level. 

\section{Model}

In order to study the LTT-pinned stripes in $d$-wave superconductors (dSC), we self-consistently solve the Bogoliubov-de Gennes (BdG) equations  on a square lattice for the mean-field Hamiltonian 

\begin{equation}
H=-\underset{ij\sigma}{\sum}t_{ij}c_{i\sigma}^{\dagger}c_{j\sigma}-\underset{i\sigma}{\sum}\mu n_{i\sigma}+U\underset{i\sigma}{\sum}(\frac{n_{i}}{2}-\sigma m_{i})n_{i\sigma}+\underset{i\delta}{\sum}(\Delta_{i\delta}c_{i\uparrow}^{\dagger}c_{i+\delta\downarrow}^{\dagger}+h.c.),
\end{equation}

where the hopping amplitudes between nearest-neighbor and next-nearest-neighbor sites \textit{i} and \textit{j} are given by $t_{ij}=t=1$ and  $t_{ij}=t'=-0.3$, respectively. The chemical potential $\mu$ is adjusted to fix the average doping level $x$. In the fourth term,  $\delta\in\left\{ \pm\hat{\mathbf{x}},\pm\hat{\mathbf{y}}\right\} $ denote unit vectors to nearest-neighbor sites. The electron number operator for spin $\sigma$ at site  \textit{i} is given by $n_{i\sigma}=c_{i\sigma}^{\dagger}c_{i\sigma}$, and the self-consistent quantities are:
\begin{eqnarray}
n_{i} & = & \left\langle n_{i\uparrow}+n_{i\downarrow}\right\rangle \\
m_{i} & = & \frac{1}{2}\left\langle n_{i\uparrow}-n_{i\downarrow}\right\rangle \\
\Delta_{i\delta} & = & -V\left\langle c_{i\uparrow}c_{i+\delta\downarrow}\right\rangle \end{eqnarray}

where $V$ is an attractive pairing interaction strength. A standard Bogoliubov transformation leads to the BdG matrix equation
\begin{equation}
\Biggl(\begin{array}{cc}
\xi_{\uparrow} & \Delta\\
\Delta^{*} & -\xi_{\downarrow}^{*}\end{array}\Biggr)\Biggl(\begin{array}{c}
u_{n}\\
v_{n}\end{array}\Biggr)=E_{n}\Biggl(\begin{array}{c}
u_{n}\\
v_{n}\end{array}\Biggr),\end{equation}
with the matrix elements defined by
\begin{eqnarray}
\xi_{\sigma}u_{i} & = & -\underset{j}{\sum}t_{ij}u_{j}+(-\mu+U(\frac{n_{i}}{2}-\sigma m_{i}))u_{i},\\
\Delta u_{i} & = & \underset{\delta}{\sum}\Delta_{i\delta}u_{i+\delta}.\end{eqnarray}
These can be solved self-consistently by iterative updates of the pairing potential $\Delta_{i\delta}$, the charge density $n_{i}$ and the local magnetization $m_{i}$ 
\begin{eqnarray}\label{nself}
n_{i} & = & 1-\frac{1}{2}\underset{n}{\sum}(\mid u_{ni}\mid^{2}-\mid v_{ni}\mid^{2})\tanh\left(\frac{E_{n}}{2kT}\right)\\
m_{i} & =- & \frac{1}{4}\underset{n}{\sum}(\mid u_{ni}\mid^{2}+\mid v_{ni}\mid^{2})\tanh\left(\frac{E_{n}}{2kT}\right)\\
\Delta_{\delta i} & = & \frac{V}{4}\underset{n}{\sum}(u_{ni}v_{ni+\delta}^{*}+v_{ni}^{*}u_{ni+\delta})\tanh\left(\frac{E_{n}}{2kT}\right)\label{Deltaself}
\end{eqnarray}
where $T$ is the temperature. The sums in Eqs.\eqref{nself}-\eqref{Deltaself} should include all eigenstates $n$.\cite{JWHarter:2007}

\section{Results and Discussion}

\begin{figure}[t!]
\includegraphics[clip=true,width=0.7\columnwidth]{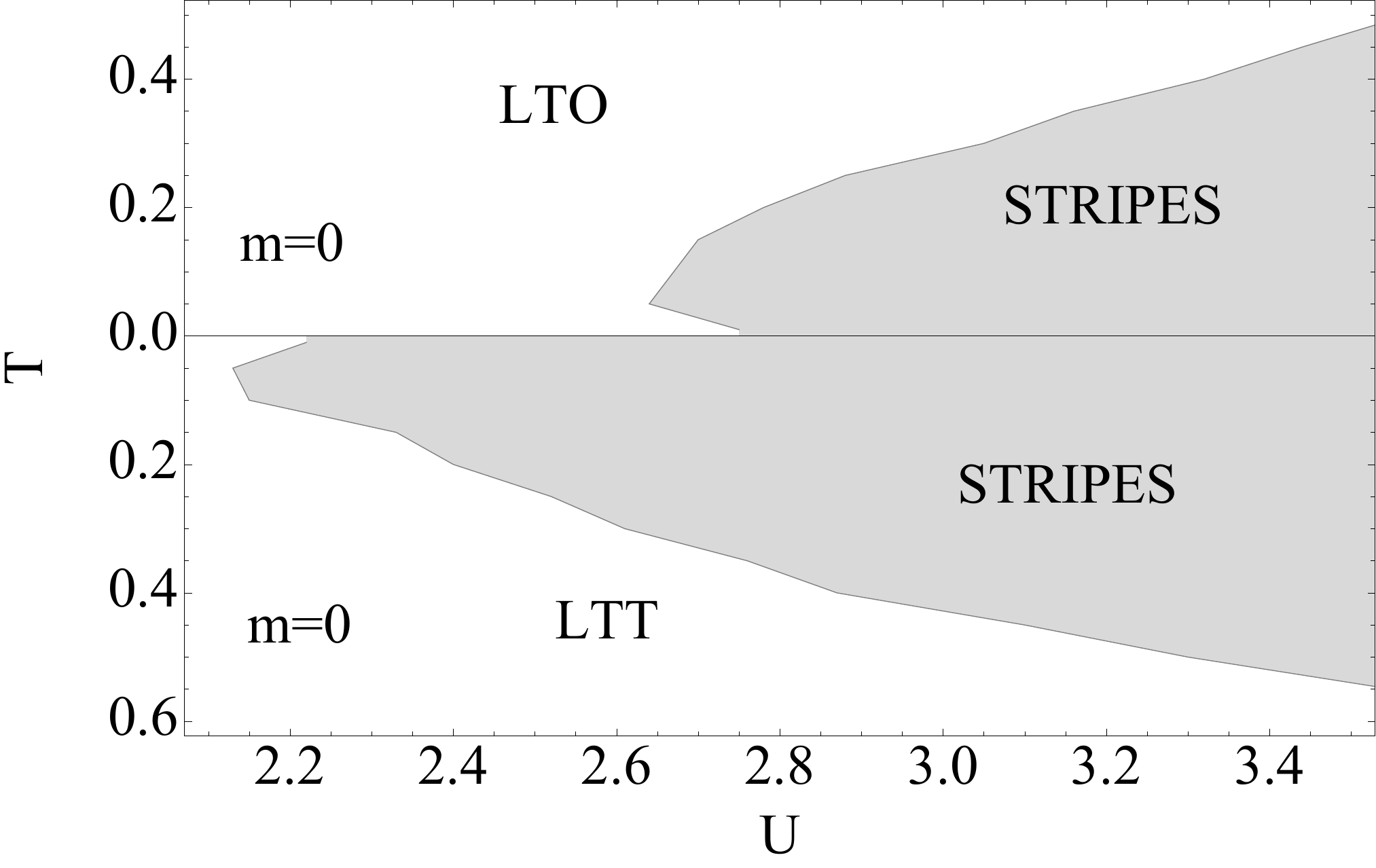}
\caption{$U\!-\!T$ phase diagram for existence of stripes comparing bulk LTT and LTO lattices.} \label{phasediagram}
\end{figure}

The LTT lattice is modeled by a hopping asymmetry $\Delta t=(t_y-t_x)/t_y$. In Fig. \ref{phasediagram} we show the $U\!-\!T$ phase diagram for bulk LTO ($\Delta t = 0$) and LTT ($\Delta t = 5\%$) phases at $x=0.10$ (and $V=1$). As seen explicitly, the critical Coulomb interaction $U_c$ for generating a stripe phase is clearly lowered by the hopping asymmetry which, in the present weak-coupling approach, is caused by a modified Stoner criterion due to the distorted Fermi surface of the LTT phase. 

\begin{figure}[b!]
\includegraphics[clip=true,height=0.4\columnwidth,width=0.4\columnwidth]{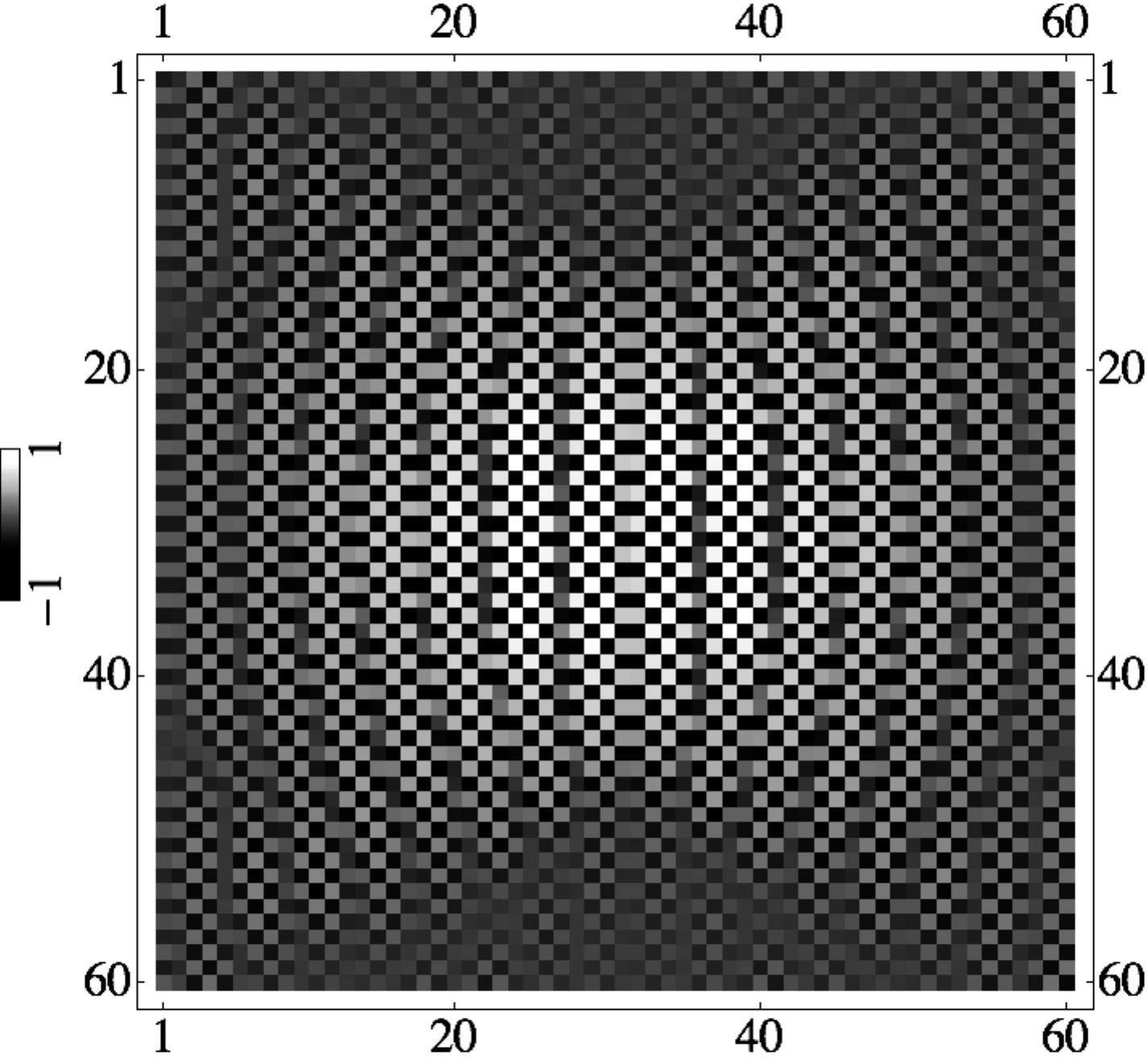}
\includegraphics[clip=true,height=0.3\columnwidth,width=0.4\columnwidth]{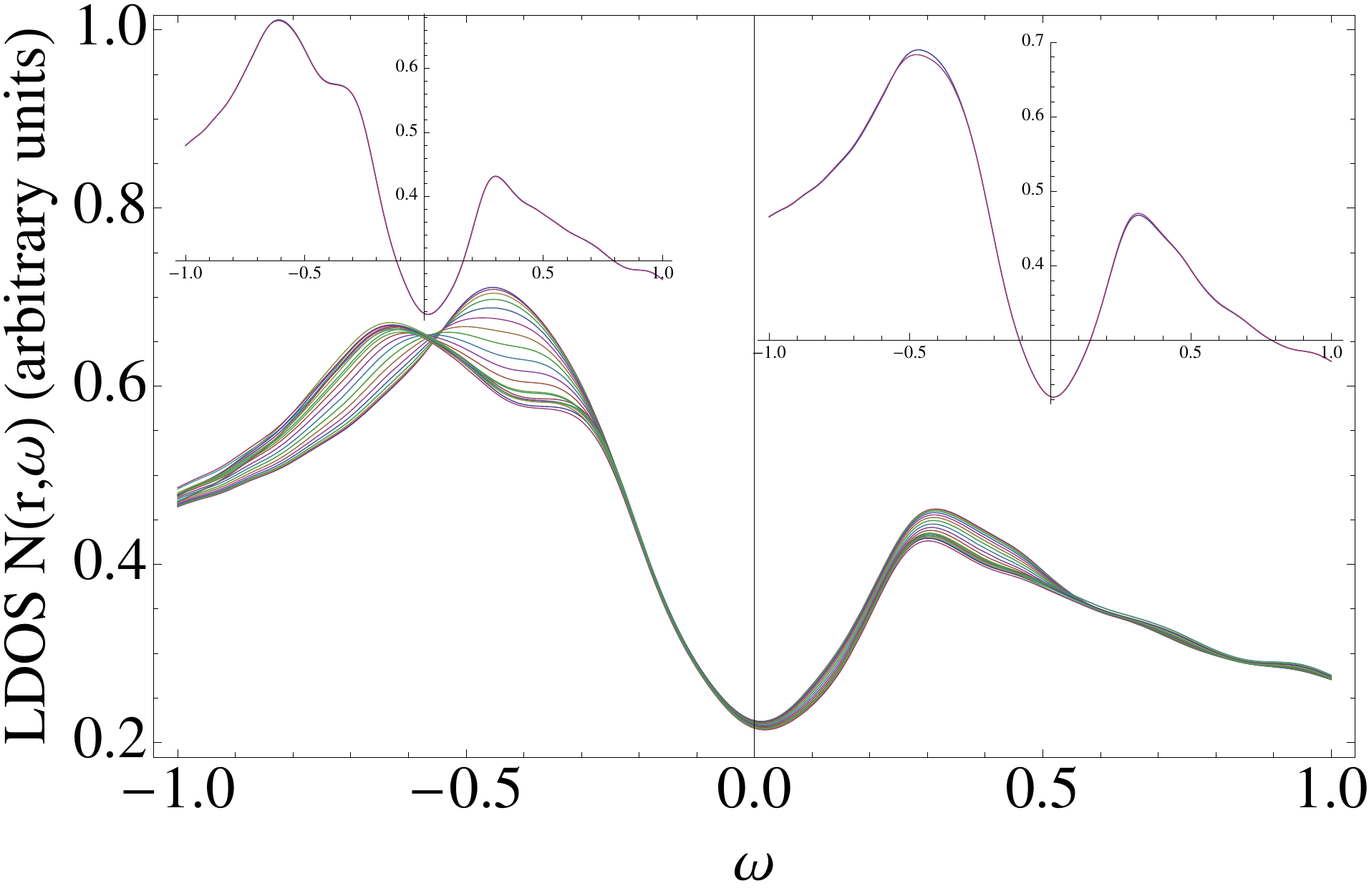}
\caption{ (left) Example of a locally LTT-pinned stripe showing the real-space magnetization in a LTO lattice with a $5\times 5$ LTT patch in the center of the system. The parameters are the same as in Fig. \ref{phasediagram} with $U=2.55$, i.e. below the critical $U_c$ for generating stripes the in LTO phase. (right) LDOS along a vertical cut [sites $(31,1) \rightarrow (31,30)$] through the center of the system shown on the left. As seen, the static stripe order of the local LTT region exhibits a clear VHS evident as a second peak near $\omega \sim -0.65$ in the left inset. By contrast, only a single peak exists at negative $\omega$ (right inset) in the purely dSC state because of a collapse of the VHS and the coherence peak.} \label{magnandLDOS}
\end{figure}

Several earlier studies have used a BdG approach with Hubbard correlations to study impurity-induced antiferromagnetism (AF) and stripes.\cite{zhu,chen04,BMAndersen06,JWHarter:2007,BMAndersen:2007,andersen10} In that case, potential scatterers generate sub-gap resonance states which can be favorably split by a local magnetization. Typically this mechanism gives rise to in-gap peaks or kinks in the LDOS even though Coulomb correlations tend to restore the pure $d$-wave gap form.\cite{BMAndersen:2006,andersen08,garg,tklee2009} In Fig. \ref{magnandLDOS} we show the resulting magnetization in a system with a $5\times 5$ LTT patch in the center of the system. The local LTT mechanism for stripe pinning does not contain impurity resonances but the induced magnetism may, depending on band parameters and $U$, cause a in-gap kink similar to coexisting phases of dSC and AF.\cite{andersen09} In the present case, however, the spatial modulations of the LDOS upon entering the LTT stripe region exist near the gap edges as seen in Fig. \ref{magnandLDOS} with a spatially dependent spectral weight shift between the coherence peak and the van Hove singularity (VHS). For the doping level used in Fig. \ref{magnandLDOS}, the LDOS inside the pinned stripe region exhibits a VHS clearly separated from the coherence peak, whereas for the pure dSC phase these two peaks merge and result in the overall appearance of slightly particle-hole asymmetric coherence peaks. In real samples such features in the LDOS will coexist with additional nano-scale gap variations arising from e.g. dopant disorder and pseudo-gap phases, and would therefore be very difficult to extract.   

\begin{figure}[]
\includegraphics[clip=true,width=0.7\columnwidth]{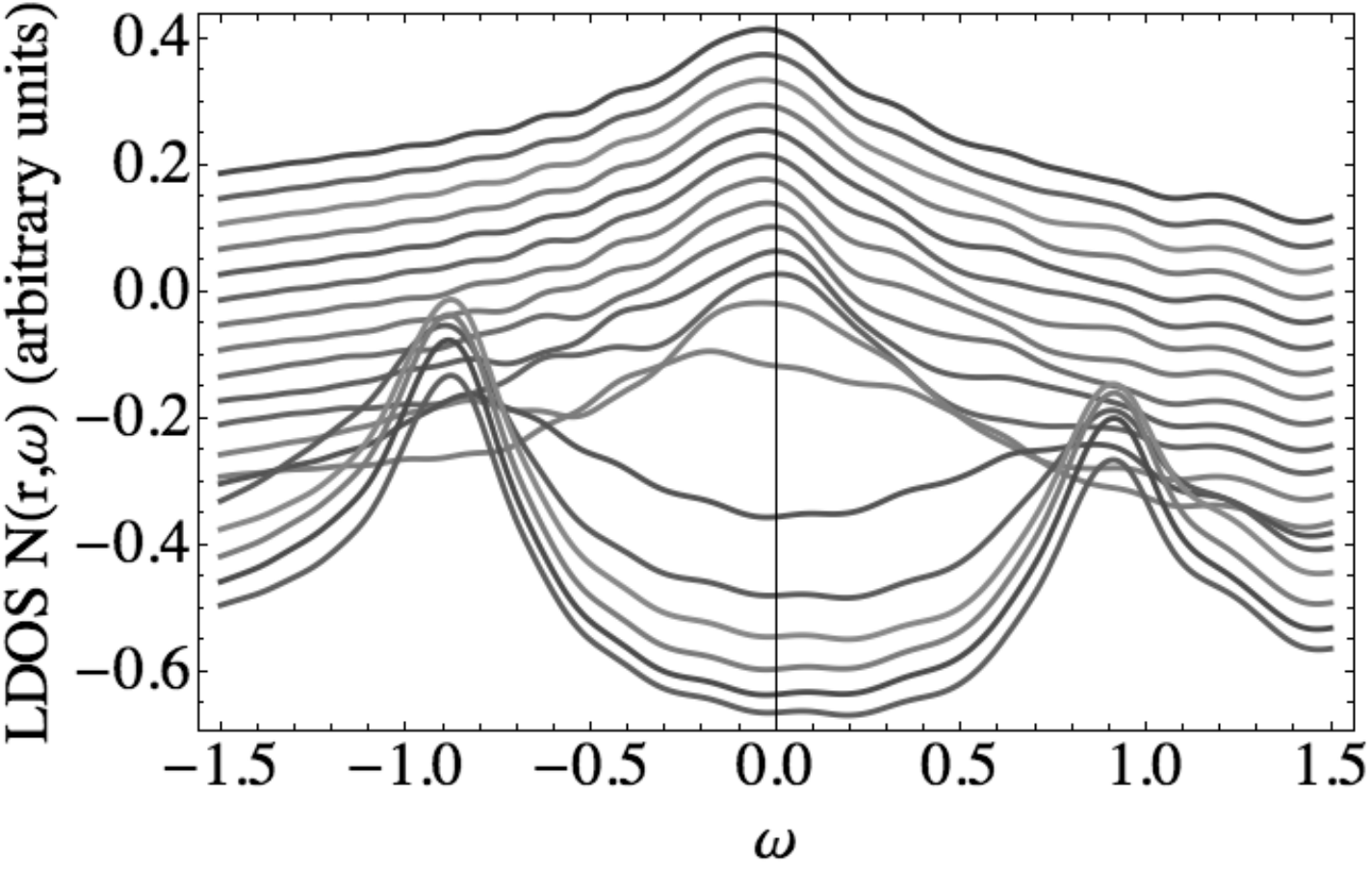}
\caption{Waterfall plot for the LDOS going from a superconducting region (bottom) into a normal region (top) in the overdoped regime. Parameters: $V=2$, $U=0$, and $x=0.20$).} \label{ldosVHS}
\end{figure}

The VHS appearing as a (logarithmic) peak in the DOS is generally not observed in the cuprates due either to disorder effects similar to Fig. \ref{magnandLDOS}, or self-energy damping or strong correlations. Recently, however, Piriou {\it et al.}\cite{piriou} detected a VHS in the overdoped regime of Bi$_2$Sr$_2$CuO$_{6+\delta}$ by STM measurements. The VHS is visible in normal regions which coexist with gapped patches at the nano-scale in the overdoped regime. The present real-space approach allows a direct simulation of such a granular normal state-superconductor surface. Figure \ref{ldosVHS} shows a waterfall plot of the evolution of the LDOS as the tip enters the normal region with a clear emergence of a VHS near the Fermi level similar to the experimental observation in Ref. \onlinecite{piriou}. 

\section{Conclusions}

We have used a BdG real-space approach to model stripe pinning by local LTT distortions, and coexisting normal state/superconducting regions revealing the long sought for VHS. For both these cases, we have calculated the associated LDOS modulations. 

\section{Acknowledgement}

B.M.A. acknowledges support from The Danish Council for Independent Research $|$ Natural Sciences.

\end{document}